\documentclass[conference]{IEEEtran}
\IEEEoverridecommandlockouts

\usepackage{cite}
\usepackage{amsmath,amssymb,amsfonts}
\usepackage{textcomp}
\usepackage{makecell}

\usepackage[utf8]{inputenc}
\usepackage{enumitem}
\usepackage{algorithmic}
\usepackage{graphicx}
\usepackage[utf8]{inputenc}
\usepackage{textcomp}
\usepackage{xcolor}
\usepackage{multirow}
\usepackage{float} 
\usepackage{tabularx}
\usepackage{algorithm}
\usepackage{adjustbox}
\usepackage{soul}
\sethlcolor{yellow}

\usepackage{bm}
\makeatletter
\AtBeginDocument{\DeclareMathVersion{bold}
\SetSymbolFont{operators}{bold}{T1}{times}{b}{n}
\SetMathAlphabet{\mathrm}{bold}{T1}{times}{b}{n}
\SetMathAlphabet{\mathit}{bold}{T1}{times}{b}{it}
\SetMathAlphabet{\mathbf}{bold}{T1}{times}{b}{n}
\SetMathAlphabet{\mathtt}{bold}{OT1}{pcr}{b}{n}
\SetSymbolFont{symbols}{bold}{OMS}{cmsy}{b}{n}
\renewcommand\boldmath{\@nomath\boldmath\mathversion{bold}}}
\makeatother
\def\BibTeX{{\rm B\kern-.05em{\sc i\kern-.025em b}\kern-.08em
    T\kern-.1667em\lower.7ex\hbox{E}\kern-.125emX}}

\begin{document}

\title{Identification of Community Structures in Networks Employing a Modified Divisive Algorithm}

\author{
Ghazal Ghajari\textsuperscript{1},
Hooshang Jazayeri-Rad\textsuperscript{2},
Mashalla Abbasi Dezfooli\textsuperscript{3} \\
\textsuperscript{1}Computer Science and Engineering, Wright State University, Ohio, USA\\
\textsuperscript{2}Department of Instrumentation and Automation, Petroleum University of Technology, Ahwaz, Iran\\
\textsuperscript{3}Department of Computer Sciences, Islamic Azad University, Khuzestan Science and Research, Ahwaz, Iran\\
\textsuperscript{1}ghajari.2@wright.edu, \textsuperscript{2}Jazayerirad@put.ac.ir, \textsuperscript{3}Abbassi@gmail.com
}

\maketitle

\begin{abstract}
In numerous networks, it is vital to identify communities consisting of closely joined groups of individuals. Such communities often reveal the role of the networks or primary properties of the individuals. In this perspective, Newman and Girvan proposed a modularity score $(Q)$ for quantifying the power of community structure and measuring the appropriateness of a division. The $Q$ function has newly become a significant standard. In this paper, the strengths of the $Q$ score and another technique known as the divisive algorithm are combined to enhance the efficiently of the identification of communities from a network. To achieve that goal, we have developed a new algorithm. The simulation results indicated that our algorithm achieved a division with a slightly higher $Q$ score against some conventional methods.
\end{abstract}


\section{Introduction}
In recent years, there has been an interest within the science communities in the properties of networks of many kinds, including the Internet, the World Wide Web, citation networks, transportation networks, software call graphs, email networks, food webs, and social and biochemical networks \cite{a6,a7,a8,a9}. One property that has attracted specific consideration is that of “community structure” in which the vertices in networks are often discovered to be clustered into tightly inter weaved groups with a high concentration of within-group edges and a lower concentration of between-group edges. 

The identification of the community structure in a network is normally designed as a procedure for converting the network into a tree (see Fig. \ref{fig2}). In the resulting tree (which is generated by our algorithm and known as the dendrogram in social sciences), the leaves are the nodes whereas the branches join nodes or (at a higher level) groups of nodes, thus giving a ranked structure of communities nested within each other. Numerous techniques to perform this conversion are known in the literature. Atypically utilized method is the so-called hierarchical clustering \cite{a10}. In this method, for every pair $i,j$ of nodes in the network, one computes a weight $W_{i,j}$, which quantifies how strictly connected their vertices are. Starting from the set of all nodes and no edges, links are iteratively considered between pairs of nodes in the course of decreasing weights. In this way nodes are accumulated into larger and larger communities, and the tree isconstructed to the root, which signifies the entire network. Algorithms of this kind are known as agglomerative. For the other category of algorithms, called divisive, the order of construction of the tree is reversed: one starts with the entire graph and iteratively eliminates the edges, thus dividing the network progressively into smaller and smaller detached sub-networks categorized as the communities. The central point in a divisive algorithm is the selection of the edges to be eliminated, which have to be those joining communities and not those inside them.

The very well-known Girvan-Newman (GN) technique \cite{a3} is a divisive technique \cite{a3}. The GN technique was first proposed in \cite{a3} and later improved in \cite{a1}. This algorithm is based on the iterative elimination of edges with high “betweenness” scores that appears to categorize such structure with some sensitivity. This algorithm has been utilized by a number of authors in the study of such different systems as: networks of email messages, social networks of animals, collaborations of jazz musicians, metabolic networks, and gene networks \cite{a1, a3, a11,a12,a13,a14,a15}. As specified by Newman and Girvan \cite{a1}, the principal disadvantage of their algorithm is the high computational loads it requires. Generally, it executes in the worst-case time of $O(m^2n)$ on a network with m edges and n vertices, or $O(n^3)$ on a sparse network (one for which m scales with n in the limit of large n, which includesbasically all networks of existing scientific interest, with the possible exception of food webs). With typical computer resources accessible at the time of writing, this limits the algorithm’s application to networks of a few thousand vertices as a maximum, and significantly less than this for interactive applications. More and more, however, there is concern in the study of much larger networks; citation and collaboration networks can comprise millions of vertices \cite{a16, a17}, for example, while the World Wide Web sums are in the billions \cite{a18}. 

Radicchi et al. proposed another divisive algorithm based on edge clustering coefficient in 2004 \cite{a2}. The implementation of the technique comprises only local variables and thus it executes much faster than the GN algorithm. In 2005, another new algorithm using optimization was proposed by Duch et al.\cite{a19}. This algorithm left behind the other algorithms in finding a higher modularity $Q$ \cite{a1}.
\begin{figure}[H]
    \centering    \includegraphics[width=1\linewidth]{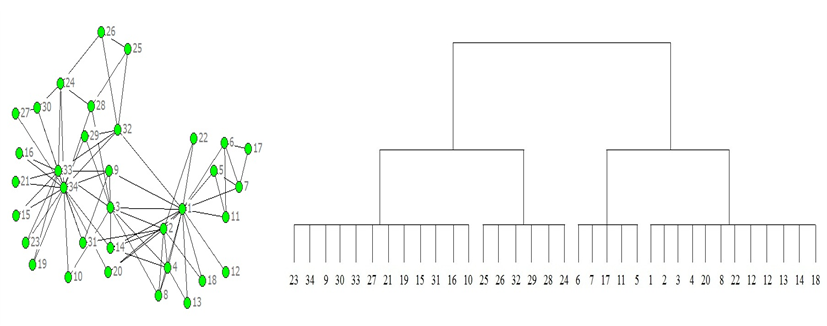}
    \caption{ A graph of a network (left) and the corresponding dendrogramgenerated by our algorithm (right) }
    \label{fig1}
\end{figure}
\section{PRELIMINARY KNOWLEDGE}
In this section, we reconsider the fundamental principles of the modularity $Q$ and two divisive algorithms; these constitute the building blocks of our proposed algorithm.
\subsection{Algorithm Based on Edge Betweenness}
The GN algorithm, which considers the edge betweenness as the weight of edges, is the leading divisive algorithm. The betweenness of an edge is defined as the number of shortest routes between pairs of nodes passing it. The algorithm to compute edge betweenness is explained in detail elsewhere \cite{a1}. Further studies show that the performance of the GN algorithm is quite acceptable when it is used in the networks in which the community structure can be straightforwardly categorized. However, in the cases where communities are hard to classify, it occasionally fails to produce a suitable result. One important reason is its tendency to divide nodes connecting to the networks lightly in the early stages, resulting in the final less reasonable community division.
\subsection{Algorithm Based on Edge Cluster Coefficient}
The algorithm suggested by Radicchi et al. \cite{a2} was initially intended for solving the efficacy problem of the GN algorithm. It presented the idea that the edge clustering factor Cij defined as below could also be used as the weight of the edges. Formally, for the edge linking node i to node j, the clustering factor is:
\begin{equation}
C_{i,j}^{(3)} = \frac{Z_{i,j}^{(3)} + 1}{\min(k_i - 1, k_j - 1)}
\label{eq:1}
\end{equation}
where $Z_{i,j}^{(3)}$ is the number of triangles comprising the edge, min $[k_i-1,k_j-1]$ is the highest possible sum of triangles and $k_i$ ($k_j$) is the degree of node $i$ (node $j$). It is worth noting that when a node links to the networks via a single edge, $C_{i,j}^{(3)}$ is infinite since the denominator is zero. This algorithm relies on the hypothesis that within communities where links are comparatively compressed, there are more triangles than those created by the edges between communities and thus the clustering factors of the edges connecting nodes within the same community is likely to be greater than that of the edges connecting the nodes of dissimilar communities.

In view of the presence of high order cycles \cite{a2} in the networks, the characterization of the edge clustering factor in a wider sense can be expressed as:
\begin{equation}
C_{i,j}^{(g)} = \frac{Z_{i,j}^{(g)} + 1}{S_{i,j}^{(g)}}
\label{eq:2}
\end{equation}
where $Z_{i,j}^{(g)}$ is the sum of cyclic structures of g order comprising that edge and $S_{i,j}^{(g)}$ is the utmost possible sum of cyclic structures. Experiments reveal that this algorithm executes much faster than the GN algorithm, making the calculation of some large networks feasible. However, its performance mainly hinges on the sum of g order cycles in a particular network. Also, when the edge clustering factors of a number of edges are equal, acommon case in the networks, the algorithm may select one edge randomly, subsequently resulting in a less reasonable division.

\subsection{Modularity $Q$}
Here, we conciselyexplain theprinciple of modularity $Q$ first presented by Newman and Girvan \cite{a1}. References can be made to \cite{a4, a20, a21} for details. Modularity is a characteristic of a network and anexplicitplanned division of a network into communities. Modularity function is a function of adefinite division, with larger magnitudes representing stronger community structure. Hence, we can basically find good divisions of a network into communities by improving the modularity over conceivable divisions.

Assuming an undirected graph $G = (V, E)$, let $A_{vw}$ be the element of nearby matrix of graph G thus: 
\begin{equation}
A_{vw} = 
\begin{cases}
1 & \text{if vertices } v \text{ and } w \text{ are connected;} \\
0 & \text{otherwise.}
\end{cases}
\label{eq:3}
\end{equation}
And assume that the vertices are divided into communities such that vertex $v$ belongs to $C_v$. Then, the $Q$ function can be expressed as follows: 
\begin{equation}
Q = \frac{1}{2|E|} \sum_{v,w} \left( A_{vw} - \frac{d_v d_w}{2|E|} \right) \delta(C_v, C_w)
\label{eq:4}
\end{equation}
where $|E|$ is the number of edges in graph $G$, $d_v$ and $d_w$ are the 
degrees of both vertices $v$ and $w$, and $\delta$ is a function which can 
be expressed as: 
\begin{equation}
\delta(C_v, C_w) = 
\begin{cases}
1 & \text{if } C_v = C_w; \\
0 & \text{otherwise.}
\end{cases}
\label{eq:5}
\end{equation}
To simplify the computation of the $Q$ function in a recursive process, we obtain the following relation: 
\begin{equation}
\begin{split}
Q = \frac{1}{2|E|} \sum_{v,w} \left( A_{vw} - \frac{d_v d_w}{2|E|} \right) \delta(C_v, C_w) \\
= \frac{1}{2|E|} \sum_{c=1}^{k} E_c - \frac{1}{4|E|^2} \sum_{c=1}^{k} D_c^2
\end{split}
\label{eq:6}
\end{equation}
where $E_c$ is twice the number of edges whose both vertices belong to the community $c$, $D_c$ is the total degree of the vertices inside the community $c$, and $k$ is the sum of communities. From the above expression, we can find:

$Q=$ (number of edges inside communities) \text{-} (predictable number of such edges).

If no edges can be found to join vertices across clusters then $Q = 1$, and on the contrary if the accumulation of inter cluster edges is no better than random then $Q = 0$, just as described in \cite{a4}. Newman and Girvan discovered that tangible un-weighted networks with high community structure generally have $Q$ scores ranging from $0.3$ to $0.7$. The modularity $Q$ may also be simplyencompassed to weighted graphs. Compared to other cost functions such as: ratio cut \cite{a22}, normalized cut \cite{a23}, and min-max cut \cite{a24}, the modularity $Q$ has two concrete benefits: (1) It can be employed to spontaneously determine the optimal number of communities; (2) It can efficiently overwhelm the bias of balanced communities.
\section{ALGORITHMS}
In this work, the  $MoveQ$  technique introduced in \cite{a25} is used for enhancing the modularity. This notion as a post processing step is applied to the already determined communities after running a heuristic algorithm on them. The technique iteratively moves the borderline’s vertices having the least fitness into other communities until a prime state with the greatest $Q$ score is satisfied.

Determination of borderlines for communities is inefficient because we have to examine each vertex by regularly searching the adjacency list and after discovering store them into a set borderline.Therefore, for improving the discovering time of borderlines divisive algorithms instead of agglomerative algorithms are employed. These algorithms identify and cut the interface edges between the communities, so two vertices at the end of such edge, in fact become the borderlines of two communities. Therefore, determination of the borderlines is included in the execution of the divisive algorithm that requires no extra time. Consequently, when an edge is being broken, its two ending vertices will be included into the borderline sets of the two communities.

Among the divisive algorithms, two of them have been considered. To identify the interface edges between communities, one of these two methods uses the Edge Betweenness measure $(EB)$ and the other employs the Edge Cluster Coefficient $(CC)$ measure. In some works \cite{a26, a27} these two measures are combined together to determine a Weight $(W)$ factor for detecting the interface edges between communities. The computation for the $EB$ measure is more time consuming than the $CC$ measure. Moreover, the necessary time for computing $W$ is essentially the same as the computation time for $EB$, which makes this approach almost impracticable for large networks. 

In this work two consecutive measures have been used. Initially, the $CC$ measure, which is faster for conversion of a large community into smaller ones, is employed. However, this measure can’t determine communities which are loosely connected to the other communities. In this approach, a loosely-coupled community will be considered as a part of a stronger community. Following the application of the $CC$ measure, the $EB$ measure is applied to the established communities. Because these communities are smaller than the original ones, using the $EB$ measure is justifiable. In addition, it is possible to discover the smaller communities which are loosely linked to larger ones. The  $MoveQ$  algorithm has been used in the execution of both of these two algorithms. 

\begin{figure}[H]
    \centering
\includegraphics[width=1\linewidth]{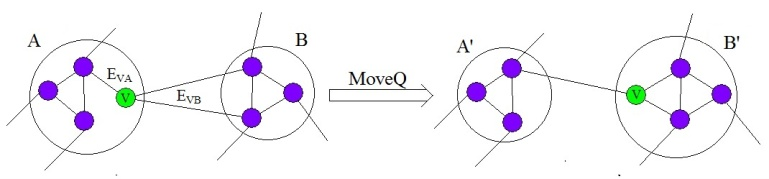}
    \caption{An example of computing the $MoveQ$ score}
    \label{fig2}
\end{figure}

The kernel of the refinement process contains computing  $MoveQ$  scores between the neighboring vertices and moving the ones with the biggest scores to the corresponding communities in each level. An example of computing the  $MoveQ$  score is shown in Fig.\ref{fig2}.
Based on Eq. \ref{eq:6}, the expression for the calculation of the $MoveQ$ score can be obtained. Considering the example above, we have:
\begin{subequations}
\renewcommand{\theequation}{7\alph{equation}}
\begin{equation}
Q_A = \frac{E_A}{2|E|} - \left( \frac{D_A}{2|E|} \right)^2,
\tag{7-a}
\label{eq:7a}
\end{equation}
\begin{equation}
Q_B = \frac{E_B}{2|E|} - \left( \frac{D_B}{2|E|} \right)^2.
\tag{7-b}
\label{eq:7b}
\end{equation}
\end{subequations}

Once the green vertex $V$ moves from $A$ to $B$, then we get: 
\begin{subequations}
\renewcommand{\theequation}{8\alph{equation}}
\begin{equation}
Q_{A'} = \frac{E_A - 2E_{VA} - E_V}{2|E|} - \left( \frac{D_A - D_V}{2|E|} \right)^2,
\tag{8-a}
\label{eq:8a}
\end{equation}
\begin{equation}
Q_{B'} = \frac{E_B - 2E_{VB} - E_V}{2|E|} - \left( \frac{D_B - D_V}{2|E|} \right)^2.
\tag{8-b}
\label{eq:8b}
\end{equation}
\end{subequations}

Therefore, the $MoveQ$ score will be given as: 

\begin{align}
\text{MoveQ} &= Q_{A'} + Q_{B'} - Q_A - Q_B \notag \\
             &= \frac{E_{VB} - E_{VA}}{|E|} + \frac{1}{2|E|^2} \left( D_A D_V - D_V^2 - D_B D_V \right)
\label{eq:9}
\end{align}
where $E_{VA}$ is the number of edges between the vertex $V$ and the community $A$, and $E_{VB}$ is the number of edges between the vertex $V$ and the community $B$.

 The $CC$ refinement (CCR) algorithm based on the edge cluster coefficient, which includes a refinement phase, can be summarized as:
\subsection{The CCR Algorithm}
Input: Given the graph $G = (V, E)$ having n vertices. Output: A set of intermediate $C$. Method: CCR$(G)$

\begin{enumerate}[label=\textit{\roman*.}]
    \item Calculate the $CC$ scores for all edges in the network.
    \item Find the edge $e_{ij}$ with the lowest score and remove it from the graph.
    \item Store $v_i$ and $v_j$ in the borderline sets of two communities.
    \item Recalculate $CC$ for all the remaining edges.
    \item Repeat from step \textit{$ii$} until two sub communities are obtained.
    \item For each community $C$
    
    For each vertex $v$ in the borderline:
    \begin{enumerate}[label=\textit{\alph*.}]
    \item Calculate the $v$’s $MoveQ$ scores, and select a community $C$ with the maximal value;
    \item If the maximum of $MoveQ$ scores $maxMoveQ > 0$ then move the $v$ into $C$;
    \item Update the borderline sets;
    \item If the iteration reaches a predefined threshold then go to Step \textit{vi}.
\end{enumerate}

    \item If the $Q$ factor does not reach the maximum value then goto Step $i$ else stop. 
\end{enumerate}

Here, we add a refinement phase into the edge betweenness $(EB)$ algorithm. In addition, we use the output generated by the previously described CCR algorithm as input to this algorithm. The resulting algorithm is then called CCR-EBR. The CCR-EBR algorithm can then be summarized as:
\subsection{The CCR-EBR Algorithm}
Input: Given a set of intermediate $C$ generated by the CCR algorithm. 
Output: A set of communities. Method: CCR-EBR$(C)$
\begin{enumerate}[label=\textit{\roman*.}]
    \item Calculate the $EB$ scores for all edges in the community.
    \item Find the edge $e_{ij}$ with the highest score and remove it from the graph.
    \item Store $v_i$ and $v_j$ in borderline sets of the corresponding communities.
    \item Recalculate $EB$ for all the remaining edges.
    \item Repeat from Step \textit{ii} until two sub communities are obtained.
    \item For each community $C$

    For each vertex $v$ in borderline: 

    \begin{enumerate}[label=\textit{\alph*.}]
    \item Calculate $v$’s MoveQ scores, and select a community $C$ with the maximum value;
    \item If $maxMoveQ > 0$ then move the $v$ into $C$;
    \item Update borderline sets;
    \item If the iteration reaches a predefined threshold then go to Step \textit{vi}.
\end{enumerate}
    \item If the $Q$ factor does not reach the maximum value then go to Step \textit{i} else stop.
\end{enumerate}

\section{EXPERIMENTAL RESULTS}
In this section, we present results ofour algorithmon several real-world and synthetic networks. We compare the quality of optimization achieved by our methods to the performance of the approaches previously employed. We present an extensive comparison of the results of our algorithms with those of past heuristics. 

We tested our methods on several networks and were able to identify community structures with very high modularity values. The networks that we used in our work are unweighted and undirected. The test networks include: a network representing the friendships between 34 members of a karate club in the US over a period of 2 years (Zachary’s Karate Club) \cite{a28}; an interaction network of the characters from Victor Hugo’s novel Les miserables (MIS) \cite{a29}; a network of books about US politics published around the time of the 2004 presidential election, sold by the online bookseller Amazon.com and compiled by V. Krebs (Political books) \cite{a30}; a network of common adjectives and nouns in the novel David Copperfield by Charles Dickens (Adjnoun) \cite{a21}; a network representing the schedule of Division I football games for the 2000 season compiled by Girvanand Newman (Football games) \cite{a6}; a collaboration network of jazz musicians (Jazz) \cite{a15}; and finally a network of email contacts between studentsand faculty (Email) \cite{a13}. A description of the real world data sets used in this work is given in Table \ref{tab1}.
\begin{table}[h]
\centering
\caption{THE REAL WORLD DATASETS USED IN THIS WORK}
\begin{tabular}{|c|c|c|}
\hline
\textbf{Network} & \textbf{Vertices} & \textbf{Edges} \\
\hline
Zachary’s Karate Club (Zach) & 34 & 78 \\ \hline
Les Miserable (Lmis)         & 77 & 254 \\ \hline
Political Books (Book)       & 105 & 441 \\ \hline
Adjnoun (Adj)                & 112 & 425 \\ \hline
Football Games (Ball)        & 115 & 613 \\ \hline
Jazz                         & 198 & 2742 \\ \hline
Email (Mail)                 & 1133 & 5451 \\
\hline
\end{tabular}
\label{tab1}
\end{table}

We compare our algorithms against past published partitioning heuristics, specifically: the edge-betweenness based GN algorithm; the bottom-up heuristic of Clauset et al. \cite{a4} (denoted by CNM); Noack and Rotta’s solutions \cite{a5} (denoted by NR); the coarsening and subdivision algorithms from Noack (denoted by Sub-Div); and finally Rotta’s paper \cite{a5} (denoted by Cors). We summarize the results in Table \ref{tab2} that contains the outcomes of our algorithms (denoted by CCR and CCR-EBR) and of the past heuristics.

As tabulated in Table \ref{tab2}, for the Zachary’s karate club application of the first step of our algorithm (i.e. the RCC algorithm) generates communities, with high modularity (Q = 0.4197), which either outperforms other algorithms or generates comparable results to other algorithms. The CCR EBR achieves a comparable result to the CCR for this data set. However, the CCR-EBR algorithm $(Q = 0.5600)$ outperforms the CCR algorithm $(Q = 0.5428)$ for the Les-miserables dataset. Similar to the previous data set, it either outperforms other algorithms or generates comparable results to the other algorithms. Other data sets in the table generate similar results.

\begin{table}[H]
\centering
\caption{MODULARITY Q SCORES COMPARISON}
\resizebox{\columnwidth}{!}{%
\begin{tabular}{|c|c|c|c|c|c|c|c|}
\hline
         & \textbf{GN}    & \textbf{CNM}   & \textbf{NR}     & \makecell{\textbf{Sub-}\\\textbf{Div}} & \textbf{Cors}  & \textbf{CCR}    & \makecell{\textbf{CCR-}\\\textbf{EBR}} \\
\hline
\textbf{Zach}     & 0.401 & -     & 0.4197 & 0.419   & 0.419 & 0.4197 & 0.4197  \\ \hline
\textbf{Lmis}     & 0.540 & 0.5006& 0.5600 & -       & -     & 0.5428 & 0.5600  \\ \hline
\textbf{Book}     & -     & 0.5019& 0.5269 & 0.3992  & 0.5269& 0.5269 & 0.5269  \\ \hline
\textbf{Adj}      & -     & 0.2934& 0.3072 & -       & -     & 0.309  & 0.309   \\ \hline
\textbf{Ball}     & 0.601 & 0.5772& 0.6002 & 0.602   & 0.605 & 0.6001 & 0.6044  \\ \hline
\textbf{Jazz}     & 0.405 & -     & -      & 0.442   & 0.440 & 0.445  & 0.445   \\ \hline
\textbf{Mail}     & 0.532 & 0.5116& 0.5774 & 0.572   & 0.556 & 0.4531 & 0.5703  \\
\hline
\end{tabular}
}
\label{tab2}
\end{table}

\begin{figure}[H]
    \centering
    \includegraphics[width=1\linewidth]{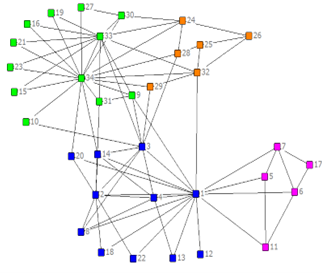}
    \caption{Division of the Zachary’s karate club by our algorithm}
    \label{fig3}
\end{figure}

As shown in Fig. \ref{fig3}, our algorithm divides the Zachary’s karate club into four communities. In Fig. 4 the division of the Les-miserables dataset using the CCR algorithm is shown. Fig. 5 shows the division of the same data set using the CCR EBR algorithm.

Our algorithm was implemented in visual Basic. Net. All the experiments are performed on a laptop with a 2.4GHz Intel\textsuperscript{\textregistered} Core\textsuperscript{\texttrademark} i5 CPU and 4GB of RAM. The operating system platform was Windows 7 home premium.
\begin{figure}[H]
    \centering
    \includegraphics[width=1\linewidth]{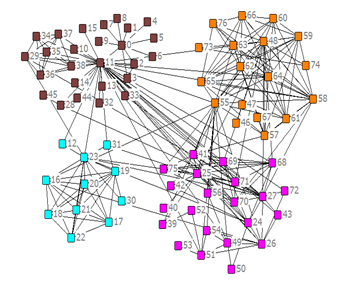}
    \caption{Division of the Les-miserables dataset by the CCR algorithm}
    \label{fig4}
\end{figure}
\begin{figure}[H]
    \centering
    \includegraphics[width=1\linewidth]{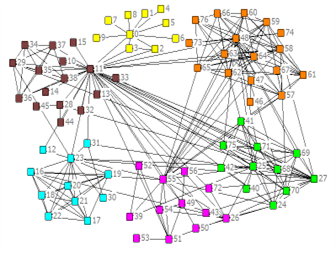}
    \caption{Division of the Les-miserables dataset by the CCR-EBR algorithm}
    \label{fig5}
\end{figure}

\section{Result}

\section{Conclusion and Future Work}
In this paper we have developed two methods known as the CCR and CCR-EBR algorithms. The CCR algorithm is based on the cluster coefficient measure which can efficiently convert the main community into smaller sub-communities. The CCR-EBR algorithm is based on the edge betweenness measure employed in the GN algorithm. It utilizes the approximate sub-communities generated by the CCR algorithm. In the CCR-EBR algorithm, the consecutive executions of the CCR and EBR properly determine community structures in complex networks. To refine the generated communities the $MoveQ$ technique is employed in the executions of both of these algorithms. A distinct feature of our technique is that the post-processing (refinement) phase used by the conventional algorithms \cite{a25} is included within the CCR and EBR algorithms. We have found that application of the refinement phase modifies the sub-communities, which are generally unbalanced (with respect to the number of nods), in a manner to generate more balanced communities (with almost equal number of nodes). Therefore, this reduces the total execution time required to process all sub-communities. For example, in a parallel processing system a processor can be allocated to each sub-community. Hence, the execution time will be a function of time required to process the sub community containing the greatest vertices. Clearly, the processing time will be minimized if balanced sub communities are utilized.

In addition, employing the modularity $Q$, we have tested our algorithm using different datasets. We have demonstrated that our algorithms either generated comparable results to the conventional algorithms or outperformed these algorithms in terms of accuracy.

Future extension of the work presented here may include: (i) prior to the application of the EB measure, parallel processing can be applied to the balanced communities generated by the CC measure; (ii) new algorithms which execute faster and are more efficient than the CC and EB measures used in this work can be developed.


\end{document}